\documentclass{aa}
\usepackage{graphicx}
\usepackage{epsfig}
\usepackage{amssymb}
\usepackage{epsf}
\usepackage{natbib}
\bibpunct{(}{)}{;}{a}{}{,}

\renewcommand\[{\begin{equation}}
\renewcommand\]{\end{equation}} 
\newcommand{\commu}{{\rm C}[\rho, \Phi]}
\newcommand{\Ep}{\dot{E}}
\newcommand{\etah}{\eta_{\rm h}}
\newcommand{\etag}{\eta_{\rm g}}

\newcommand{\HI}{\ion{H}{i}}
\newcommand{\Md}{M_{\rm d}}

\newcommand{\mH}{m_{\rm p}}
\newcommand{\pp}{P}
\newcommand{\Rd}{R_{\rm d}}

\newcommand{\Rh}{\tilde{R}_{\rm h}}
\newcommand{\Rot}{\tilde{R}_{0}}
\newcommand{\Rhom}{\tilde{R}_{m}}
\newcommand{\zti}{\tilde{z}}
\newcommand{\Rti}{\tilde{R}}
\newcommand{\Rg}{R_{\rm g}}

\newcommand{\rhonda}{\rho_{\rm 0h}}
\newcommand{\rhong}{\rho_{0}}

\newcommand{\phieff}{\Phi_{\rm eff}}

\newcommand{\phitot}{\Phi_{\rm tot}}
\newcommand{\qg}{q_{\rm g}^2} 
\newcommand{\qh}{q_{\Phi}^2}
\newcommand{\hg}{h_{\rm g}}

\newcommand{\rhoe}{\rho_{\rm e}}
\newcommand{\rhoh}{\rho_{\rm h}}

\newcommand{\rhope}{\rho_{\rm e}(\Phi_{\rm eff})}
\newcommand{\tcool}{t_{\rm cool}}
\newcommand{\tdyn}{t_{\rm dyn}}

\newcommand{\tdrag}{t_{\rm drag}}
\newcommand{\rc}{r_{\rm c}}
\newcommand{\va}{v_0^2}
\newcommand{\vel}{v_{\varphi}}
\newcommand{\velq}{v_{\varphi}^2}

\newcommand{\velqI}{v_{\varphi, 1}^2}
\newcommand{\velqII}{v_{\varphi, 2}^2}
\newcommand{\Lt}{L_{\rm tot}}
\newcommand{\SB}{\Sigma}

\newcommand{\Ri}{R_i}
\newcommand{\Rii}{R_{i+1}}

\newcommand{\Rmax}{R_{\rm{max}}}
\newcommand{\Rmin}{R_{\rm{min}}}

\newcommand{\zj}{z_j}
\newcommand{\zjj}{z_{j+1}}

\newcommand{\zmax}{z_{\rm{max}}}
\newcommand{\zmin}{z_{\rm{min}}}

\newcommand{\rhoij}{\rho^i_j}
\newcommand{\phiRij}{{_R\Phi^i_j}}
\newcommand{\phizij}{{_z\Phi^i_j}}
\newcommand{\ppij}{\pp^i_j}

\newcommand{\ppRij}{_R\ppij}
\newcommand{\delRi}{\Delta R_i}
\newcommand{\delzj}{\Delta z_j}
\newcommand{\epa}{\Ep^i_j}
\newcommand{\epb}{\Ep^{i+1}_j}
\newcommand{\epc}{\Ep^i_{j+1}}
\newcommand{\epd}{\Ep^{i+1}_{j+1}}

\begin{document}

   \title{Hydrostatic models for the rotation of extra-planar gas\\
          in disk galaxies}

    \author{M. Barnab\`e\inst{1,2},
            L. Ciotti\inst{2},
            F. Fraternali\inst{3}
            \and
            R. Sancisi\inst{1,4}}

    \offprints{M. Barnab\`e}

   \institute{
     Kapteyn Astronomical Institute, University of Groningen, P.O. Box 800, 
     9700 AV Groningen, The Netherlands\\
     e-mail: \texttt{M.Barnabe@astro.rug.nl}
     \and
     Dipartimento di Astronomia, Universit\`a di Bologna,
     via Ranzani 1, I-40127 Bologna, Italy
     \and    
     Theoretical Physics, University of Oxford, 1 Keble Road, 
     Oxford, OX1 3NP, UK
     \and                      
     INAF-Osservatorio Astronomico di Bologna, via Ranzani 1, 
     I-40127 Bologna, Italy
   }

   \date{Submitted May 9, 2005, revised August 17, 2005}

   \abstract{We show that fluid stationary models are able to
   reproduce the observed, negative vertical gradient of the rotation
   velocity of the extra-planar gas in spiral galaxies.  We have
   constructed models based on the simple condition that the pressure
   of the medium does not depend on density alone (baroclinic instead
   of barotropic solutions: isodensity and isothermal surfaces do not
   coincide).  As an illustration, we have successfully applied our
   method to reproduce the observed velocity gradient of the lagging
   gaseous halo of NGC 891. The fluid stationary models discussed here
   can describe a hot homogeneous medium as well as a ``gas'' made of
   discrete, cold \HI~clouds with an isotropic velocity dispersion
   distribution. Although the method presented here generates a
   density and velocity field consistent with observational
   constraints, the stability of these configurations remains an open
   question.

   }

   \titlerunning{Hydrostatic models for the rotation of extra-planar gas}
   \authorrunning{Barnab\`e et al.}
   \maketitle
\section{Introduction}
\label{intro}

Observations at various wavelengths show that some spiral galaxies are
surrounded by a gaseous halo. This extra-planar gas is multiphase: it
is detected in \HI~\citep*[e.g.,][]{Swat.97}, H$\alpha$
\citep{Rand.00, Ross.04}, and X-ray observations \citep{Wang.01,
Stri.04}.  In particular, high-sensitivity \HI~observations of edge-on
galaxies like NGC 891 \citep{Swat.97, Frat.04a} and UGC 7321
\citep{Matt.03} reveal neutral gas emission up to large distances from
the plane (e.g. see the \HI~map of NGC~891 in \citeauthor{Frat.05}
\citeyear{Frat.05}) and the presence of a negative vertical gradient
in the gas rotational velocity.  A similar decrease of the rotational
velocity with distance from the plane has also been observed in the
diffuse ionized gas halo of NGC 891 \citep{Rand.97} and of NGC 5775
\citep{Rand.00}.

The two major issues regarding the extra-planar gas are those of its
origin and of its dynamical state. These are strictly related. For
example, the halo gas could be the result of cosmological accretion
\citep[e.g., see][]{Binn.05}, or of a galactic fountain
\citep{Shap.76, Breg.80}, or of both.  Different structures and
kinematics would be expected for these cases.  Thus, the study of the
origin and dynamics of the extra-planar gas not only is important in
itself, but may also provide a new insight on the formation and the
structure of spiral galaxies.  Furthermore, it may bring new evidence
on the vertical distribution of dark matter.  In this paper we focus
on the problem of the dynamical state of the extra-planar gas.

Two ``extreme'' types of models have been considered for the
extra-planar gas and in particular for its decreasing rotational
velocity: the \emph{ballistic} and the \emph{fluid homogeneous}
models.

Ballistic models describe the gas as an inhomogeneous collection of
clouds, subject only to the gravitational potential of the galaxy: for
example, in the galactic fountain model ionized gas is ejected from
the galactic disk due to stellar winds and supernova explosions, and
then cools and falls back ballistically \citep{Breg.80}.  These models
are able to explain vertical motions of the cold (\HI) and warm
(H$\alpha$) gas components observed in several spiral galaxies
\citep[e.g.][]{Frat.04b, Boom.05}.  However, \citet*{Coll.02} have
tried a ballistic model to describe the extra-planar ionised gas of
NGC\,891 and found problems in reproducing the observed kinematics.
These authors suggest that the discrepancy could be solved by
considering ``the presence of drag between disk and halo, such as
through magnetic tension or viscous interactions between
clouds. Alternatively, an outwardly directed pressure gradient could
explain the gas kinematics''. 

We note here that the concept of pressure in the physics of the
interstellar medium is a complex one, as there are several
contributing sources of pressure, e.g. thermal, kinetic (or
turbulent), magnetic, cosmic ray and radiation \citep[e.g., see][~and
references therein]{Boul.90}. For simplicity, in the following we will
restrict our discussion to thermal and kinetic pressure.
In fluid homogeneous models, the extra-planar gas is described as a
stationary rotating fluid without any motion along the radial and
vertical directions, with the galaxy gravitational field balanced by
the pressure gradient and the centrifugal force. Until now, this
approach has not been fully explored in all its possibilities, and
only a few attempts have been made \citep[e.g., see][]{Benj.02}.  Here
we extend a preliminary analysis of fluid homogeneous stationary
models \citep{Barn.05} to explain the vertical gradient of rotational
velocity observed in the extra-planar gas.  In particular, we show
that models in which the gas pressure does not depend on the density
alone (\emph{baroclinic} configurations) are able to reproduce the
observed vertical gradient.  In addition, we show that baroclinic
solutions could provide the drag invoked by \citet{Coll.02}.  This
suggests that a correct description of the extra-planar gas dynamics
may be found in ``hybrid'' ballistic-fluid stationary models.
Finally, we address the question of the physical interpretation of our
solutions and suggest, as an alternative to the hypothesis of a hot
homogeneous medium, the possibility of a ``gas'' of cold \HI~clouds
described by the stationary (fluid) Jeans equations, with the
sustaining pressure given by a globally isotropic velocity dispersion
tensor.

The paper is organized as follows: in Sect.~\ref{fluid.app}
and~\ref{modelli}, we briefly introduce the baroclinic solutions and
derive simple and general rules for their construction.  In
Sect.~\ref{ngc891} we construct a fluid homogeneous and stationary
model for the galaxy NGC 891, and in Sect.~\ref{concl} we discuss the
results from an astrophysical point of view.  In the Appendices a
simple, fully analytical model of a gas distribution with low rotation
at high~$z$ is presented, together with the numerical code adopted for
the case of NGC~891.

\section{The fluid approach}
\label{fluid.app}

\subsection{The fluid equations}
\label{fluid.eq}
 
We consider a gaseous axisymmetric system in permanent rotation, under
the influence of an axisymmetric gravitational potential
$\phitot(R,z)$: because of the axial symmetry, all the physical
variables depend only on the cylindrical coordinates $R$ and $z$.  The
stationary hydrodynamical equations for the gas are then
\[
\left\{
\begin{array}{lcl}
\displaystyle \frac{1}{\rho} \, \frac{\partial \pp}{\partial z} & = & 
\displaystyle - \frac{\partial \phitot}{\partial z} \: ,\\
&&\\
\displaystyle \frac{1}{\rho} \, \frac{\partial \pp}{\partial R} & = & 
\displaystyle - \frac{\partial \phitot}{\partial R} + \Omega^2 R ,
\end{array}
\right.
\label{eq.hydro}
\]
where $\rho$, $\pp$ and $\Omega$ denote the gas density, pressure and
angular velocity, respectively; the gas rotational velocity is given
by $\vel=\Omega R$, while $v_R =v_z =0$.  Note that $\phitot$
represents the total gravitational potential, including the gas
contribution. Later on, we will assume the condition that the gas is
not self-gravitating and therefore $\phitot =\Phi$, where the galaxy
gravitational potential $\Phi$ is the sum of the dark halo and the
stellar disk potentials.

In standard applications, as for example the setting up of initial
conditions for hydrodynamical simulations, the above equations are
solved adopting a barotropic pressure distribution and neglecting the
gas self-gravity. Thus, one fixes the gravitational potential $\Phi$
and a specific function $\pp(\rho)$, and integrates the first of
Eqs.~(\ref{eq.hydro}) with the boundary $\rho(R,0)$ or imposing
$\rho(R,\infty)=0$. The angular velocity $\Omega$ is obtained from the
second of Eqs.~(\ref{eq.hydro}). This leads to \emph{cylindrical
rotation}, i.e.  $\Omega =\Omega(R)$. In fact, according to the
Poincar\'e-Wavre theorem \citep{Lebo.67, Tass.80}, cylindrical
rotation is equivalent to the fact that the acceleration field at the
r.h.s. of Eqs.~(\ref{eq.hydro}) can be obtained from an effective
potential $\phieff$ (see Eq.~[\ref{phieff}]), or that the gas density
and pressure are stratified on $\phieff$, and so the fluid is
barotropic.

Cylindrical rotation is in disagreement with the observed vertical
gradient of the extra-planar gas rotation velocity, and this would
seem to argue against the applicability of fluid stationary models.
However, in the next Section we will show that it is possible to
construct baroclinic equilibrium solutions with a negative velocity
gradient along $z$.  Note that baroclinic solutions have been studied
in the past for problems ranging from geophysics to the theory of
sunspots and to galactic dynamics\footnote{In particular, isotropic
axisymmetric galaxy models can be interpreted as baroclinic fluid
configurations, and show streaming velocities often decreasing with
$z$ (e.g., see \citeauthor{Lanz.03}~\citeyear{Lanz.03}, and
\citeauthor{Ciot.05}~\citeyear{Ciot.05} for simple examples).}
\citep[e.g., see][~and references therein]{Ross.26, Tass.80, Waxm.78}.

\subsection{Baroclinic solutions}
\label{baroclinic}

For simplicity we restrict ourselves to non self-gravitating gas
distributions, even though several results hold also in the
self-gravitating case.  We fix the total gravitational potential
$\Phi(R,z)$ for the galaxy but, at variance with the standard
treatment, we prescribe a density distribution $\rho(R,z)$ for the gas
vanishing at infinity. (We do not address here the much more
difficult problem of a consistent assignement of non-zero pressure as
a boundary condition.). The first of Eqs.~(\ref{eq.hydro}) is
integrated for the pressure as
\[ 
\pp(R,z)=\int_z^{\infty}\rho {\partial\Phi\over \partial z'} \, {\rm d} z',
\label{pressione}
\]
where we also assume $\pp (R,\infty)=0$.  In general, the obtained
pressure $\pp$ (and the corresponding temperature $T =
\mu \mH \pp / k \rho$) can not be expressed as a function of $\rho$
only, and so the system is baroclinic.  Accordingly, the rotational
velocity field 
\[ 
\velq(R,z)={R\over\rho}{\partial\pp\over\partial R} + 
           R{\partial\Phi\over\partial R}
\label{velocita}
\]
depends both on $R$ and $z$.  The major problem posed by the
construction of baroclinic solutions is the fact that, for an
arbitrary choice of $\rho$ and $\Phi$, the existence of
\emph{physically acceptable solutions} (i.e. configurations for which
$\velq\geq 0$ everywhere) is not guaranteed.  In fact, due to the
arbitrariness of the chosen density field, a negative radial pressure
gradient in Eq. (\ref{velocita}) can be dominant for some values of
$R$ and $z$. Thus, before addressing the specific case of the
extra-planar gas in NGC~891, in the next Section we present a few
general results of a mathematical nature that will be used as
guidelines in the construction of physically acceptable baroclinic
solutions, while in Appendix~\ref{toymodel} we present a fully
analytical, baroclinic toy-model as an example of the procedure
described above.

\section{Simple families of solutions}
\label{modelli}

The starting point of the following analysis is obtained by combining
Eqs.~(\ref{pressione}) and (\ref{velocita}) and integrating by parts
with the assumption that $\pp$ and $\rho\partial\Phi/\partial R$
vanish for $z=\infty$, thus obtaining the exact relation
\[ 
{\rho\velq\over R}=\int_z^{\infty}\left( 
                   {\partial\rho\over\partial R}
                   {\partial\Phi\over\partial z'}- 
		   {\partial\rho\over\partial z'} 
		   {\partial\Phi\over\partial R} 
                   \right)\, {\rm d}z'\equiv \commu .
\label{commutatore}
\]

This ``commutator-like'' relation is not new (e.g., see
\citeauthor{Ross.26} \citeyear{Ross.26}, \citeauthor{Waxm.78}
\citeyear{Waxm.78} and, in the context of stellar dynamics,
\citeauthor{Hunt.77} \citeyear{Hunt.77}); here we note that the
positivity of the integrand in Eq.~(\ref{commutatore}) is a sufficient
(but not necessary) condition to obtain $\velq \geq 0$
everywhere.  Therefore, physically acceptable solutions are obtained
if one assume a potential $\Phi$ for which $\partial\Phi/\partial
R\geq 0$ and $\partial\Phi/\partial z\geq 0$ (the usual situation) and
a density distribution so that $\partial\rho/\partial z\leq 0$ and
$\partial\rho/\partial R\geq 0$.

In addition, the bilinearity of $\commu$ can be used to construct more
complicate solutions starting from simple, physically acceptable
``building block'' configurations. In fact, from
Eq.~(\ref{commutatore}) it follows that the rotational velocity
associated with $\rho =\rho_1 +\rho_2$ is
\[ 
\velq ={\rho_1\velqI +\rho_2\velqII\over\rho_1 +\rho_2},
\label{vel.birho}
\]
where $\rho_1\velqI/R\equiv\rm{C}[\rho_1, \Phi]$ and $\rho_2
\velqII/R\equiv \rm{C}[\rho_2, \Phi]$. Also, if $\Phi =\Phi_1 +\Phi_2$, then
\[ 
\velq =\velqI +\velqII .
\label{vel.biphi}
\]

\subsection{Gas density distributions with a factor stratified 
            on the effective potential}
\label{factor.rho}

We now elaborate in more detail the general results of
Eqs.~(\ref{commutatore})-(\ref{vel.biphi}).  Let us consider the
factorized gas density distribution
\[ 
\rho (R,z) = h(R,z)\rhope,
\label{rho(eff)}
\]
where $h$ is a non negative function and
\[ 
\phieff\equiv\Phi -\int_{R_0}^R\Omega^2(R')R' \, {\rm d}R'
\label{phieff}
\]
is the effective potential associated with the total potential $\Phi$ and
with a prescribed cylindrical rotation law $\Omega(R)$; $R_{0}$ is an
arbitrary but fixed radius.  We assume that $\rhoe$ in
Eq.~(\ref{rho(eff)}) is a solution of the equation
\[ 
\nabla\pp =-\rho\nabla\phieff
\label{hydroeq.eff}
\]
with assigned $\pp=\pp(\rho)$. For $\Omega = 0$ one obtains a
hydrostatic (and therefore barotropic) solution in the potential
$\Phi$ (that we indicate with $\rhoh$), while for $\Omega(R)\neq 0$
one has a cylindrical rotation (barotropic) solution.

In practice, by adopting factorization~(\ref{rho(eff)}) one modifies a
gas distribution with a cylindrical velocity field: this approach is
of obvious interest because families of $\rhope$ can be easily
constructed (see Appendix~\ref{buildrho}). Substituting 
Eq.~(\ref{rho(eff)}) in Eq.~(\ref{commutatore}) gives
\begin{eqnarray}
{\rho\velq\over R}&=&\int_z^{\infty}\left(
                  {\partial h\over\partial R}{\partial\Phi\over\partial z'} - 
		  {\partial h\over\partial z'}{\partial\Phi\over\partial R} 
                                    \right)\rhoe \, {\rm d}z' -\nonumber\\
		&& R\Omega^2(R)\int_z^{\infty} 
		   h{\partial\rhoe\over\partial z'} \, {\rm d}z'.
\label{vel:h.rhope}
\end{eqnarray}

Equations (\ref{vel.birho}) and (\ref{vel:h.rhope}) prove the
existence of physically acceptable baroclinic solutions. For example,
$\rho =\rhoh(\Phi)+\rhope$ leads to
\[
\velq = {\rhoe R^2\Omega^2(R)\over \rhoh +\rhoe}. 
\label{vel:eff+hyd}
\]

Another case of physically acceptable solutions is $\rho
=A(R)\rhoh(\Phi)$, with $A(R)$ increasing and approaching a constant
value for $R\gg 1$. In this case $\velq$ decreases reaching systemic
(zero) velocity at infinity\footnote{E.g., the function
$A(R)=R/(1+R)$.}.

We now apply the method we have just described to a couple of more specific 
astrophysically relevant cases: a homeoidally stratified potential and a 
razor-thin uniform disk. For simplicity, we assume in
Eq.~(\ref{rho(eff)}) a hydrostatic density factor $\rhoh(\Phi)$.

\subsubsection{Homeoidal potential}
\label{homeo+phi}

Let $\Phi(\ell)$ be an homeoidally stratified potential with $\ell^2
\equiv R^2 + z^2/\qh$ and $0 < q_{\Phi} \leq 1$; two well-known
examples are the \citet{Binn.81} logarithmic potential, and
\citet{Evan.94} spheroidal potentials. We write Eq.~(\ref{rho(eff)}) as
\[ 
\rho (R,z) = A(R) B(m) \rhoh (\Phi),
\label{h(m)rhop}
\]
where $m^2 \equiv R^2 + z^2/\qg$ and $0 < q_{\rm g} \leq 1$; $A(R)$
and $B(m)$ are positive functions. From Eq.~(\ref{vel:h.rhope}) we
have
\begin{eqnarray} 
{\rho\velq\over R}&=&\left({1\over\qh}-{1\over\qg}\right) A(R) R
                     \int_z^{\infty}{B'(m)\over m}\rhoh(\Phi) 
                     \Phi'(\ell){z'\over\ell}\, {\rm d}z'+\nonumber\\
                  & &{A'(R)\over\qh}\int_z^{\infty} 
                     B(m)\rhoh(\Phi)\Phi'(\ell){z'\over\ell}\,{\rm d}z',
\label{comm.homeo2}
\end{eqnarray}
and so $\velq\geq 0$ if
\[
\Phi'(\ell)\geq 0,\quad A'(R)\geq 0, \quad B'(m)\leq 0, \quad q_{\rm
g}\leq q_{\Phi}.
\label{cond.qg}
\]

Note that $\rhoh$ does not enter in the sufficient condition
(\ref{cond.qg}). The condition on flattenings ($q_{\rm g}\leq
q_{\Phi}$) requires that the gas density distribution must be
stratified on homeoids which are \emph{ flatter} than the isopotential
surfaces. This is, therefore, always satisfied for a flat gas
distribution in the spherically symmetric monopole-dominated far field
of any finite mass system (e.g. a stellar disk).

\subsubsection{Razor-thin uniform disk}
\label{potdisk}

We now discuss the case of a razor-thin uniform disk. What are the
conditions for having physically acceptable ($\velq >0$) solutions
near the disk?  We explore this issue by assuming $\rho
=A(R)\rhoh(\Phi)$ in Eq.~(\ref{rho(eff)}) (or $B=1$ in
Eq.~[\ref{h(m)rhop}]) and
\[ 
\Phi =2\pi G\Sigma_0 z .
\label{disk.unif}
\]

From Eq.~(\ref{vel:h.rhope}) (or Eq.~[\ref{comm.homeo2}])
\[ 
{\rhoh\velq\over R}=2\pi G\Sigma_0 {A'(R)\over A(R)} 
      \int_z^{\infty} \rhoh(\Phi) \, {\rm d}z' ,
\label{comm.disk.unif}
\]
i.e. $A'(R)\geq 0$ is the necessary and sufficient condition to have
$\velq\geq 0$ in such a case.

The physical reason for this condition is very simple.  A gas
distribution $\rho(R,z)$ not stratified on $\Phi$ as given by
Eq.~(\ref{disk.unif}) must be rotating and according to
Eq.~(\ref{velocita}) its pressure must be radially increasing. From
Eqs. (\ref{disk.unif}) and (\ref{pressione}) it follows that in the
present case the pressure is proportional to the gas column
density. This means that, in a vertical gravitational field,
$\velq\geq 0$ whenever the column density is radially increasing.
Note that this trend is consistent with the radial \HI~density
distribution observed in several spiral galaxies
\citep[see][]{Caya.94}. Obviously, the case presented by more
realistic disks (as the exponential disk in Sect.~\ref{ngc891}) will
require the explicit construction of the whole equilibrium solution to
check for the positivity of $\velq$.

\section{Application to NGC 891}
\label{ngc891}

We now apply the general results of Sect.~\ref{modelli} to the
modeling of the extra-planar gas of NGC~891 in an attempt to reproduce
its major features and in particular the decline of the rotational
velocity with increasing $z$. The rotational velocity field resulting
from this modeling will be compared with the \HI~rotation curves
derived by \citet{Frat.05}.

\subsection{The galaxy model}
\label{gal.model}

We consider a very idealized mass model for NGC~891, consisting of
three components: an exponential stellar disk, a dark matter halo with
an asymptotically flat rotation curve, and a centrally peaked density
distribution.  This model provides the gravitational potential
supporting a (non self-gravitating) baroclinic gas distribution.  Its
parameters are fixed to reproduce the observed rotation curve in the
galactic plane and to obey other observational constraints specified
in the following.

The surface density of the stellar disk is
\[ 
\Sigma(R) = {\Md\over 2\pi\Rd^2}\, e^{-R/\Rd} ,
\label{disco.exp.2}
\]
where $\Md$ and $\Rd$ are the disk mass and scale-length,
respectively. Its gravitational potential is given by
\[ 
\Phi_{\rm d}(R,z)=-{G \Md\over\Rd}\,\int_0^{\infty} 
                   {J_0(k\Rti) \,e^{-k|\zti|}\over (1+k^2)^{3/2}} \, {\rm d}k,
\label{Phi.exp}
\]
where $J_{0}$ is the zeroth-order Bessel function of the first kind,
and $\Rti\equiv R/\Rd$, $\zti\equiv z/\Rd$ \citep[e.g.][]{Binn.87}.
The observed rotation curve is reproduced by adding a
two-component mass distribution whose potential is
\begin{eqnarray} 
\Phi_{\rm h}(R,z) & = & \frac{\va}{2} \, \ln \left(\Rh^2 + \Rti^2 + 
                  \frac{\zti^2}{\qh} \right) + \nonumber\\
            &   & \frac{G M_0}{r_0} \ln \left(\frac{r}{r_0 + r} \right).
\label{Phi.log.ver} 
\end{eqnarray}

\begin{figure}[!t]
\resizebox{\hsize}{!}{\includegraphics{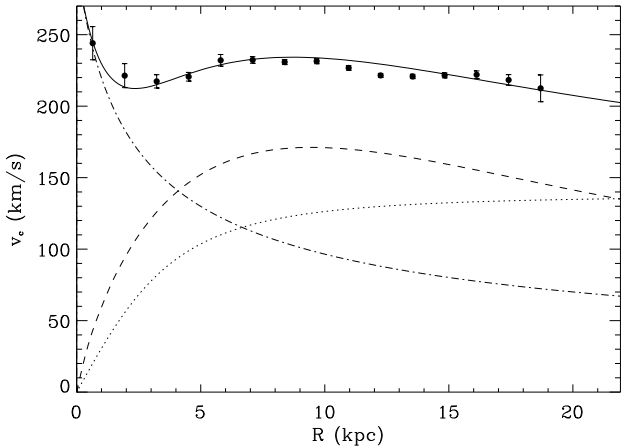}}
\caption{The adopted mass decomposition for NGC~891. Solid dots
represent the observed \HI~rotation curve \citep{Frat.05}, and the
solid line the model rotation curve. The dashed, dotted, and
dot-dashed lines show the separate contributions of the disk, the dark
matter halo, and of the spherical Jaffe mass component, respectively.}
\label{rotcur}
\end{figure}

The first component represents a \citet{Binn.81} logarithmic dark
matter halo with asymptotic velocity $v_0$ and $\Rh\equiv R_{\rm
h}/\Rd$; the corresponding potential belongs to the family of
homeoidal potentials considered in Sect.~\ref{homeo+phi}.  The second
component, needed to reproduce the steep rising of the rotation curve
in the innermost galaxy regions (see Fig.~\ref{rotcur}), is a
\citet{Jaff.83} spherically symmetric density distribution of total
mass $M_0$ and scale-length $r_0$.

The adopted values for the disk scale-length and central surface
brightness ($\Rd=4.40$ kpc, $\mu_{\rm 0 B}=$21.4 mag arcsec$^{-2}$)
were taken from \citet{Shaw.89}.  We find that the observed rotation
curve of NGC\,891 is well reproduced for $v_0=138$ km s$^{-1}$,
$R_{\rm h}=\Rd$, $q_{\Phi}=0.71$, $M_0=2.40\times 10^{10}M_{\odot}$,
$r_0=1.15$ kpc, and $M_{\rm d}=7.7\times 10^{10}M_{\odot}$. The
resulting disk B-band mass-to-light ratio is $\simeq 3.5$ (in solar
units).

\subsection{The gas distribution and its rotational velocity field}
\label{gas.distr}

Having fixed the galaxy potential well, we now turn to the choice of
the gas density distribution. Following the arguments presented in
Sect.~\ref{modelli}, we adopt the trial function 
\[ 
\rho(R,z)={\rhong\Rhom^{\beta}\over\Rot^{\alpha}}
          {(\Rot +\Rti)^{\alpha}\over (\Rhom^2+m^2)^{\beta/2}} 
	  e^{-\zti/\hg},
\label{rho.gas}
\]
where $m^2\equiv\Rti^2 +\zti^2/\qg$, $\Rhom\equiv R_{\rm m}/\Rd$,
$\Rot\equiv R_0/\Rd$, $\rho_0$ is the central gas, and $\beta >\alpha
+2$ so that the total gas mass converges. Since the logarithmic halo
belongs to the family of homeoidal potentials, $\rho$ is a factorized
distribution similar to that of Eq.~(\ref{h(m)rhop}) with an
homeoidally stratified component (flattening $q_{\rm g} < q_{\Phi}$),
a cylindrical radially increasing factor ($\alpha >0$) and an
isothermal distribution stratified on the gravity field of a
razor-thin uniform disk.

\begin{figure*}
\centering
\includegraphics[width=17cm]{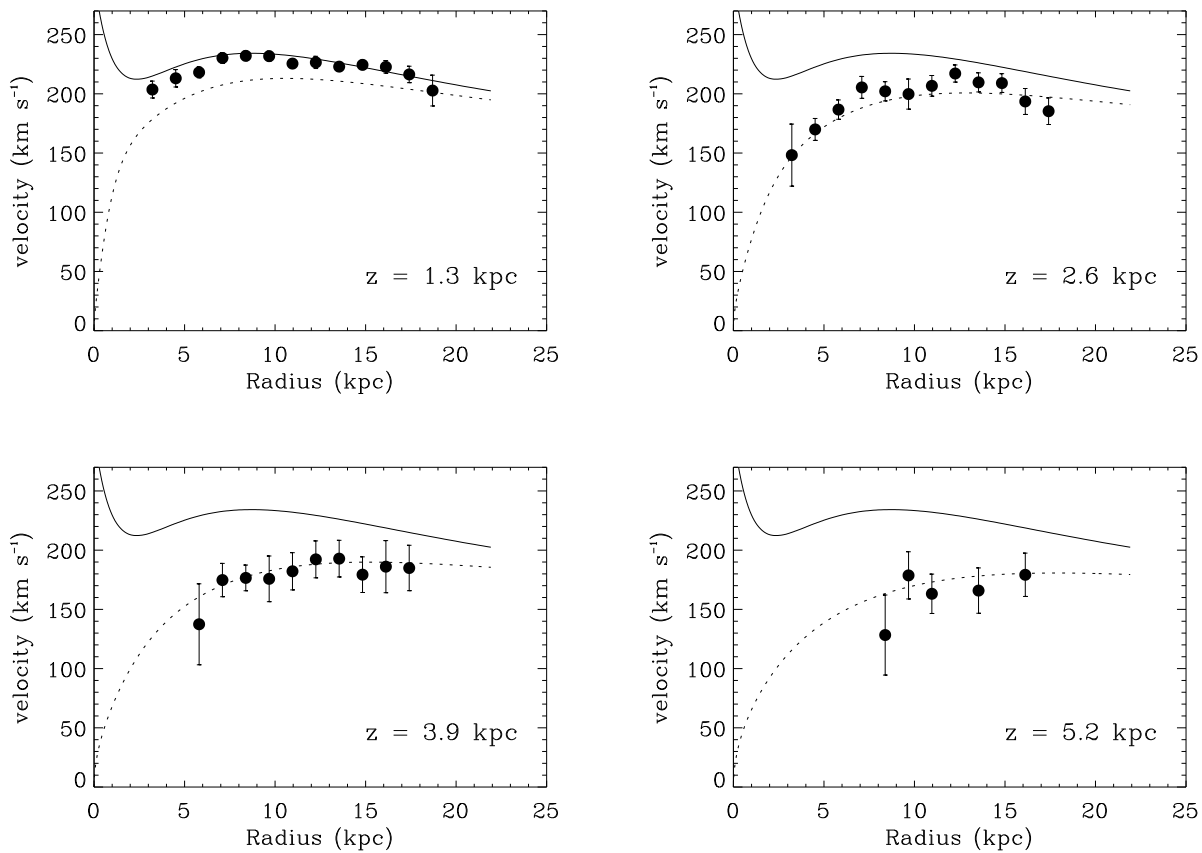}
\caption{Extra-planar \HI~rotation curves for NGC~891 (dots with error
bars) at various heights over the galactic plane
\citep[from][]{Frat.05}.  The solid curve is the rotation curve in the
galactic plane ($z=0$) as obtained from the model (see also
Fig.~\ref{rotcur}). The dotted lines are the predicted baroclinic
extra-planar rotation curves of the Reference Model.}
\label{rotmodels}
\end{figure*}

We now solve Eqs.~(\ref{eq.hydro}) for the present model. In
particular, Eq.~(\ref{pressione}) is integrated numerically with a
finite-difference scheme (Appendix~\ref{codice}), obtaining the gas
pressure, velocity and temperature values on the numerical grid.

The first result is that a large number of exploratory models (but not
all of them), in which the value of the parameters in
Eq.~(\ref{rho.gas}) is arbitrarily chosen, produce physically
acceptable rotational velocities. This confirms the validity of the
preparatory analysis in Sect.~\ref{modelli}. For example, models
without the radially increasing factor (obtained by setting $\alpha
=0$ in Eq.~[\ref{rho.gas}]), invariably have $\vel^2 <0$ in the
central regions, while models less flat than the halo potential
($q_{\rm g}>q_{\Phi}$) turn out to be non-physical in the far
field. We conclude that the ``core'' properties of the physically
acceptable models are the gas radial depression in the central regions
and a gas distribution flatter than the halo potential at large
distances. The remaining parameters play only a minor role.

We now present and discuss a specific, physically acceptable model,
which we refer to as the ``Reference Model'' (RM).  The free
parameters of the RM have been fixed to reproduce the observed
\HI~rotation curve of NGC~891 at $z=2.6$ kpc \citep{Frat.05} and the
resulting rotation curve is shown in Fig.~\ref{rotmodels} (upper right
panel, dotted line). We have obtained an acceptable curve by adopting
$\alpha =1$, $\beta =3.5$, $\Rot =1$, $\Rhom =2.25$, $\hg =1.5$ and
$q_{\rm g}=0.1$.  Note that, due to the assumption of a negligible
self-gravity for the gas, from Eqs.~(\ref{pressione}) and
(\ref{velocita}) it follows that $\vel$ and the temperature are
independent of the specific choice of $\rhong$.  Other arguments will
be used to fix the value of the latter (see Sect.~\ref{viability}).

\begin{figure}[!thb] 
\resizebox{\hsize}{!}{\includegraphics{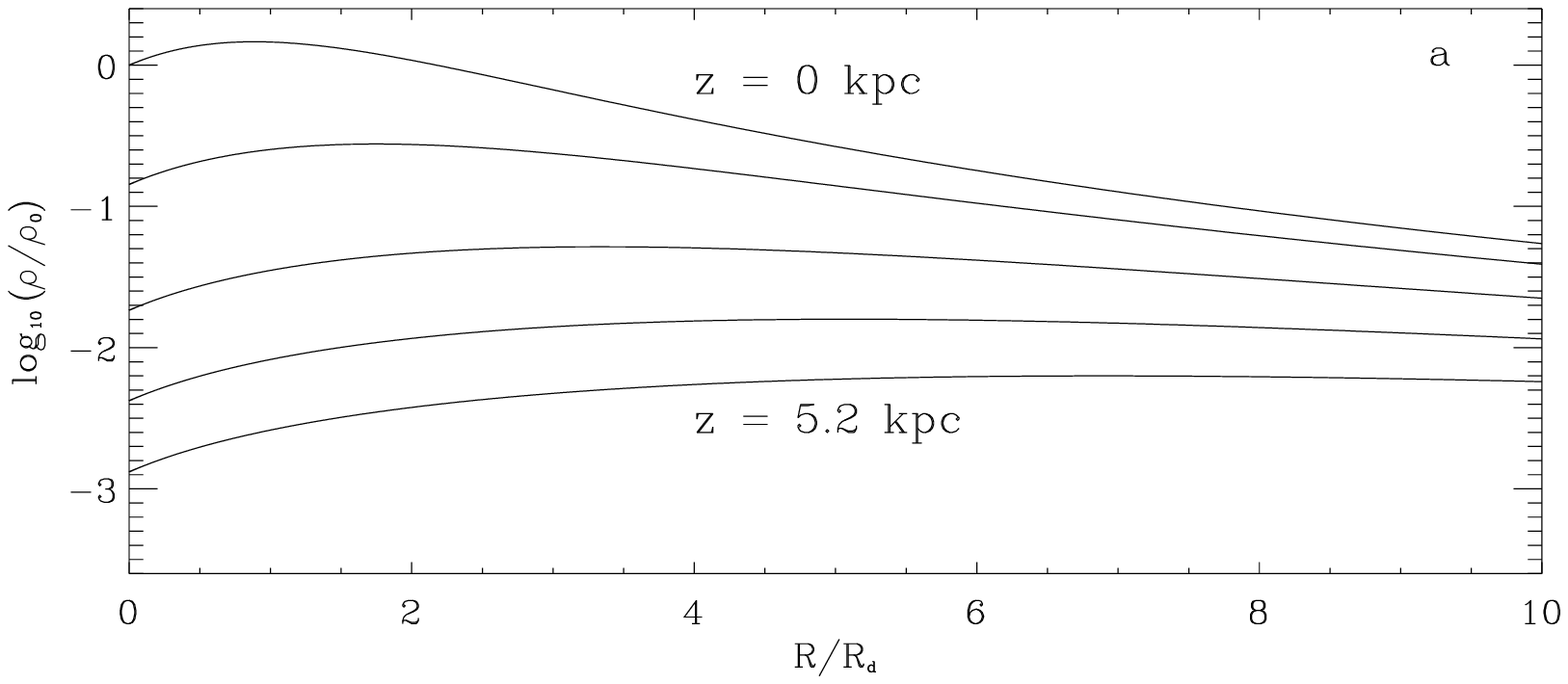}}
\vspace{0cm}
\resizebox{\hsize}{!}{\includegraphics{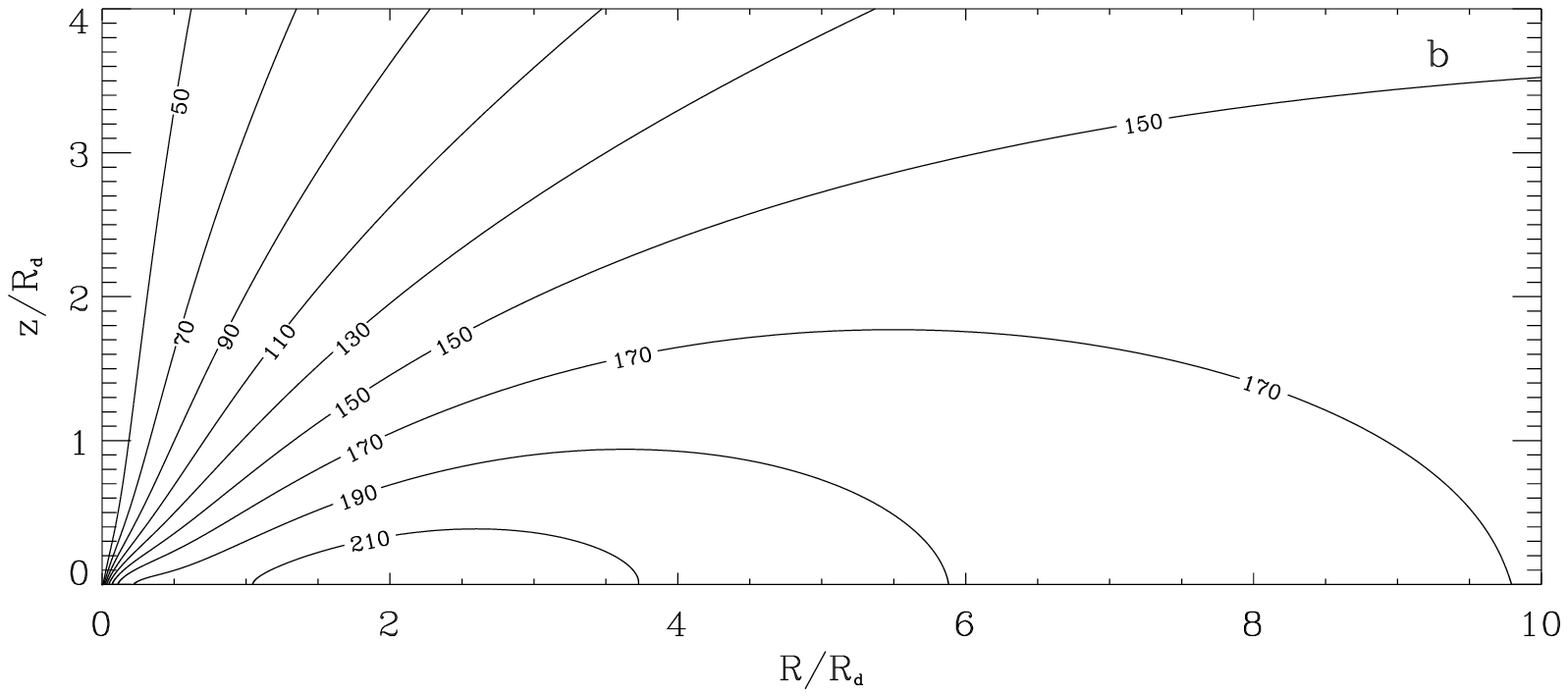}}
\vspace{0cm}
\resizebox{\hsize}{!}{\includegraphics{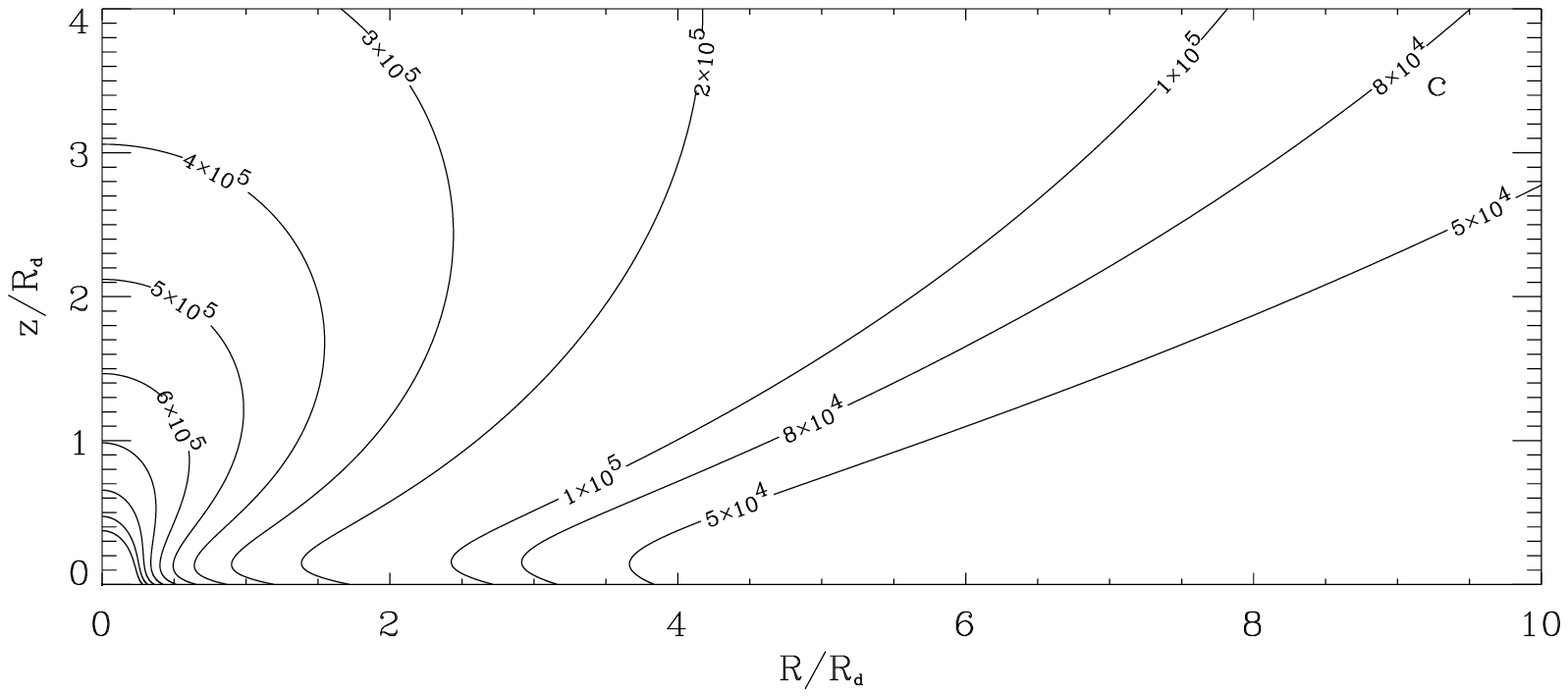}}
\caption{\emph{Panel~a:} radial profile of the density distribution
for the Reference Model (RM) at $z=0$, 1.3, 2.6, 3.9 and 5.2 kpc (same
heights as in Fig.~\ref{rotmodels}). \emph{Panels~b,c:} RM isorotation
curves $\vel = const$ (in km s$^{-1}$) and isothermal contours (in K)
in the meridional plane.}
\label{isorot}
\end{figure}

In Fig.~\ref{isorot} we show the radial profiles of the RM gas density
distribution at 5 different heights above the galactic plane
(Fig.~\ref{isorot}a) and the meridional sections of its isorotational
(Fig.~\ref{isorot}b) and isothermal (Fig.~\ref{isorot}c) surfaces.
>From Fig.~\ref{isorot}b it can be clearly seen that the rotation
velocity decreases with~$z$ at a fixed $R$, while Fig.~\ref{isorot}c
shows that the gas is warm, with temperatures ranging from $\sim 10^4$
to $\sim 10^6$~K. The hotter gas is near the $R=0$ axis, in
correspondence with the density decrease near the galaxy center, while
the lowest gas temperatures are attained near the galactic plane.

In Fig.~\ref{rotmodels} we show the observed \HI~rotation curves at
four different heights above the galactic plane ($z=1.3$, $2.6$,
$3.9$, and $5.2$ kpc, dots), and the rotation curves from the RM
(dotted lines), corresponding to horizontal sections of
Fig.~\ref{isorot}b.  The RM gas parameters have been tuned to obtain a
good fit of the observed rotation curve at $z = 2.6$ kpc. This choice
of parameters produces good fits for the curves at higher $z$ too. It
is remarkable that the gas density distribution in
Eq.~(\ref{rho.gas}), mainly built on theoretical arguments, should
lead to a predicted rotational velocity decrease for the extra-planar
gas which is so close to the observations. As to the discrepancy
between data and model predictions at $z = 1.3$ kpc, it could be real
(see Sect.~\ref{viability}), but it is possible that the observed data
are affected by the limited angular resolution of the observations
(HPBW $=28''\approx 1.3$ kpc). Details on the derivation of the
rotation curves from the observations will be presented in a
forthcoming paper (Fraternali, in preparation).

How do the RM properties depend on the specific choice of the gas
density distribution?  Among the parameters, the largest effect is
produced by the oblateness $q_{\rm g}$ and the vertical scale-height
$\hg$.  Increasing $q_{\rm g}$ has a major effect: isorotation curves
would be heavily modified for $z\gtrsim\Rd/2$, becoming almost
vertical (cylindrical rotation) when $q_{\rm g}\simeq q_{\Phi}$, which
is the limiting flattening according to condition (14). We note that a
similar behaviour is also shown by the analytical toy-model presented
in Appendix~\ref{toymodel}, even though this is based on a
substantially simpler gas and galaxy model.  A decrease of $\hg$ has a
strong effect for $z\gtrsim\Rd$, producing a cylindrical rotation
pattern extending out to $R\simeq 5\Rd$.

Variations of $\alpha$, $\beta$ and $\Rot$ produce minor effects, at
least in the regions covered by observations ($R\lesssim 5\Rd$ and
$z\lesssim 1.3\Rd$).  In particular, $\vel$ increases for increasing
$\alpha$ and decreasing $\Rot$, but the same overall structure of the
RM is kept, and the differences are hardly detectable for $z\gtrsim
1.3\Rd$. An increase of $\beta$ has an effect very similar to
increasing $\alpha$, and again there is a corresponding behaviour in
the analytical toy-model of Appendix~\ref{toymodel}.  Finally, an
increase of $q_{\Phi}$ (i.e., the adoption of a rounder dark matter
halo) at fixed $q_{\rm g}$ has only a very marginal effect on
$\vel$. We conclude that baroclinic solutions are sensitive to the
adopted gas distributions, but only marginally to the halo
flattening. It seems, therefore, that according to the fluid
stationary interpretation presented here the kinematics of the
extra-planar gas can not be used as a sensitive diagnostic tool to
investigate the flattening of dark matter haloes.

\subsection{Astrophysical interpretation}
\label{viability}

The gas kinematics from the Reference Model is remarkably similar to
that observed for the extra-planar \HI; yet, the model gas temperature
is much higher than that of \HI~(Fig.~\ref{isorot}c).  Does this mean
that fluid solutions should be abandoned as unphysical?

The gas density distribution in Eq.~(\ref{rho.gas}) may be interpreted
not only as that of a smooth, homogeneous fluid (for which pressure
and temperature are the usual thermodynamical quantities), but also as
the fluid description of the \HI~clouds distribution.  In this paper
we follow the first interpretation; however, a brief discussion of the
second interpretation, in which Eqs.~(\ref{eq.hydro}) are interpreted
as the Jeans equations for a system with isotropic velocity
dispersion, is given in Sect.~\ref{int:clouds} with further comments in
Sect.~\ref{concl}.

\subsubsection{A homogeneous gas distribution}
\label{int:gas}

Equilibrium configurations (baroclinic or not) of a homogeneous
extra-planar gas are expected to be hotter than the \HI~because of the
needed vertical pressure support against the galaxy gravitational
field. Therefore, the observed cloudy extra-planar \HI~gas can not be
described \emph{directly} with fluid homogeneous models. We note,
however, that baroclinic solutions as presented above are comparable
to (or better than) ballistic models at reproducing the observed
vertical gradients in the \HI~and H$\alpha$ rotational
velocities. Furthermore, there is, in addition to cold \HI, also hot
gas in the halo of spiral galaxies as observed in some edge on
systems, including NGC 891 \citep{Breg.94, Stri.04}.  It is therefore
important to explore the possibility that the observed \HI~traces the
kinematics of an underlying, homogeneous, hot gas.

The origin and the coexistence of hot and cold gas phases and their
kinematical coupling are central issues in the discussion of possible
realistic models for the extra-planar gas. An important question in
this respect is how long it would take for the hot gas to cool if not
resupplied with fresh energy, or to drag \HI~clouds (accreted or
produced by galactic fountains) in regular motion around the galaxy
axis. Clearly, hydrodynamical numerical simulations are needed to
fully address it.  Here, we limit ourselves to a first order analysis
of the main astrophysical aspects and implications of the Reference
Model presented above.

\begin{figure}[!tb]
\resizebox{\hsize}{!}{\includegraphics{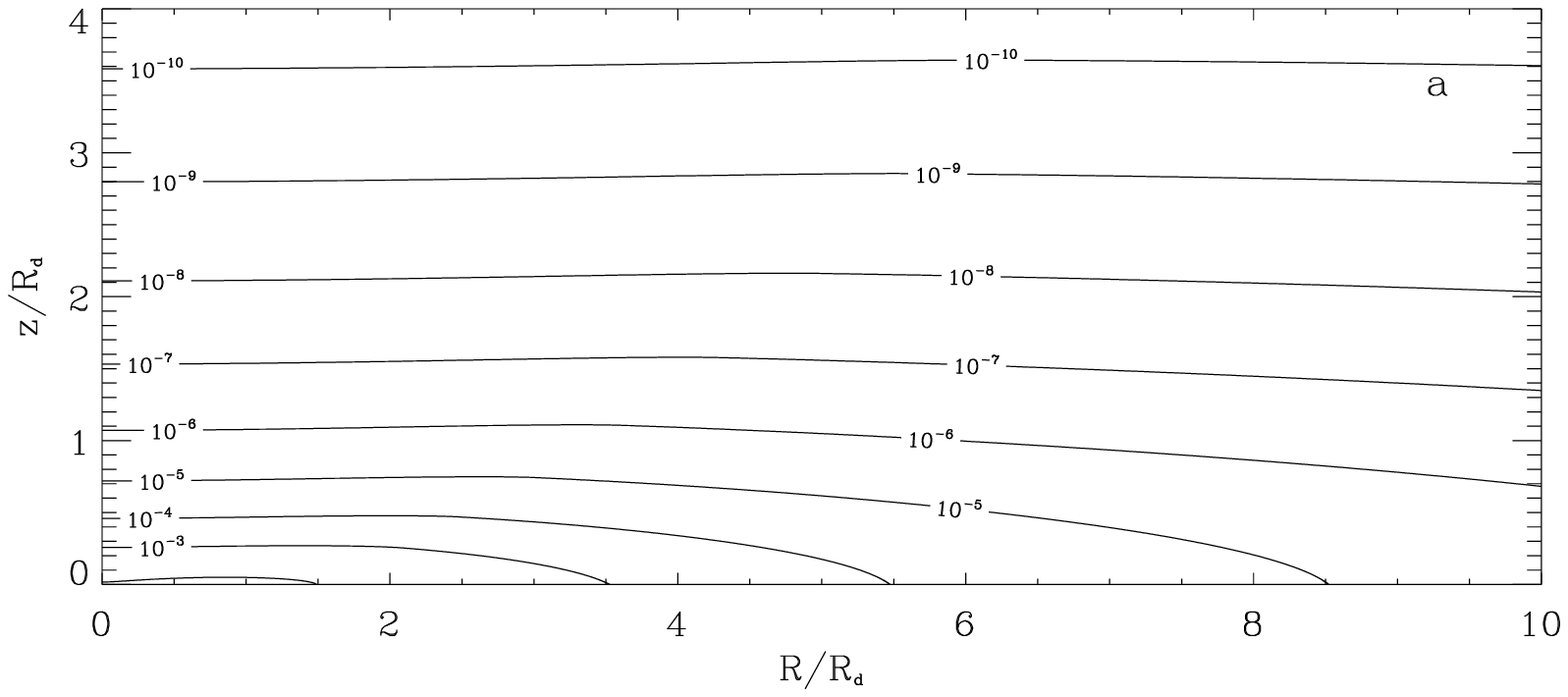}}
\vspace{0cm}
\resizebox{\hsize}{!}{\includegraphics{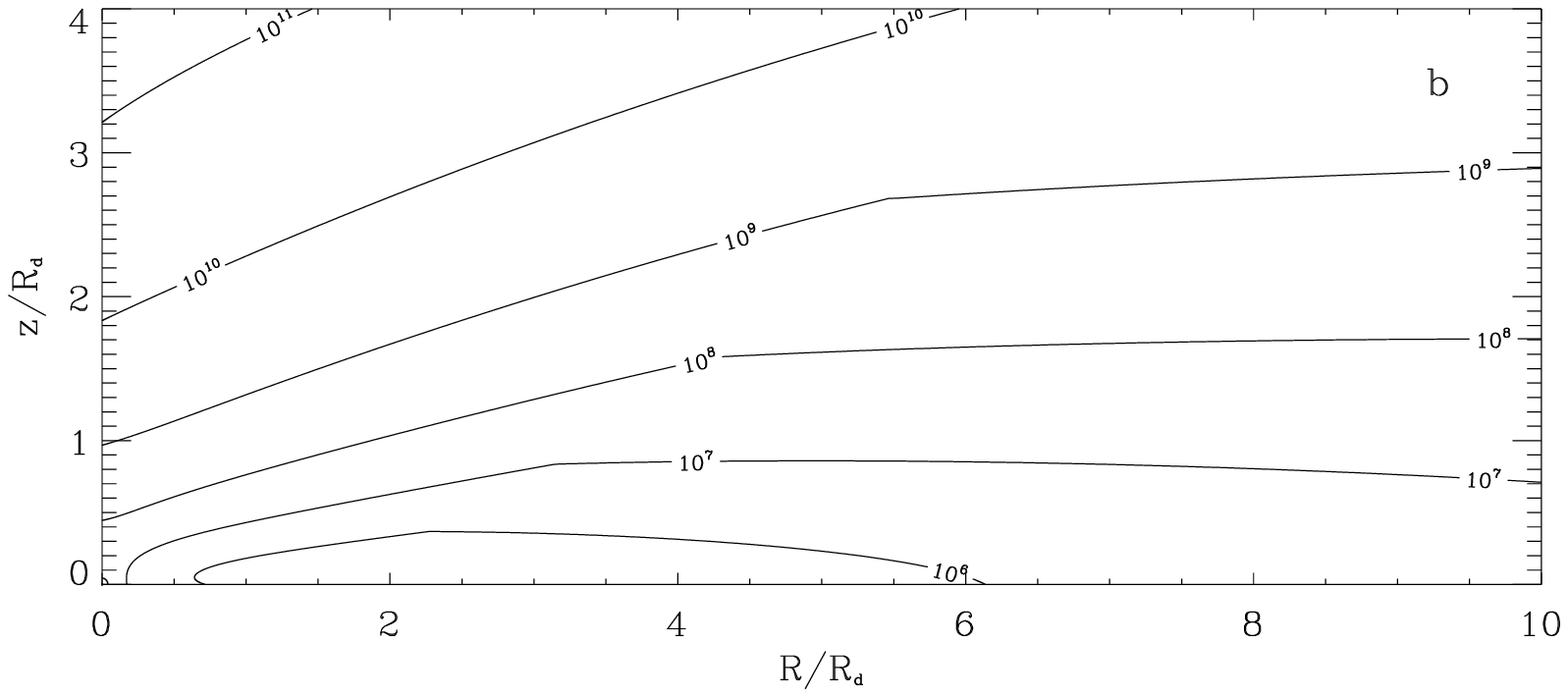}}
\vspace{0cm}
\resizebox{\hsize}{!}{\includegraphics{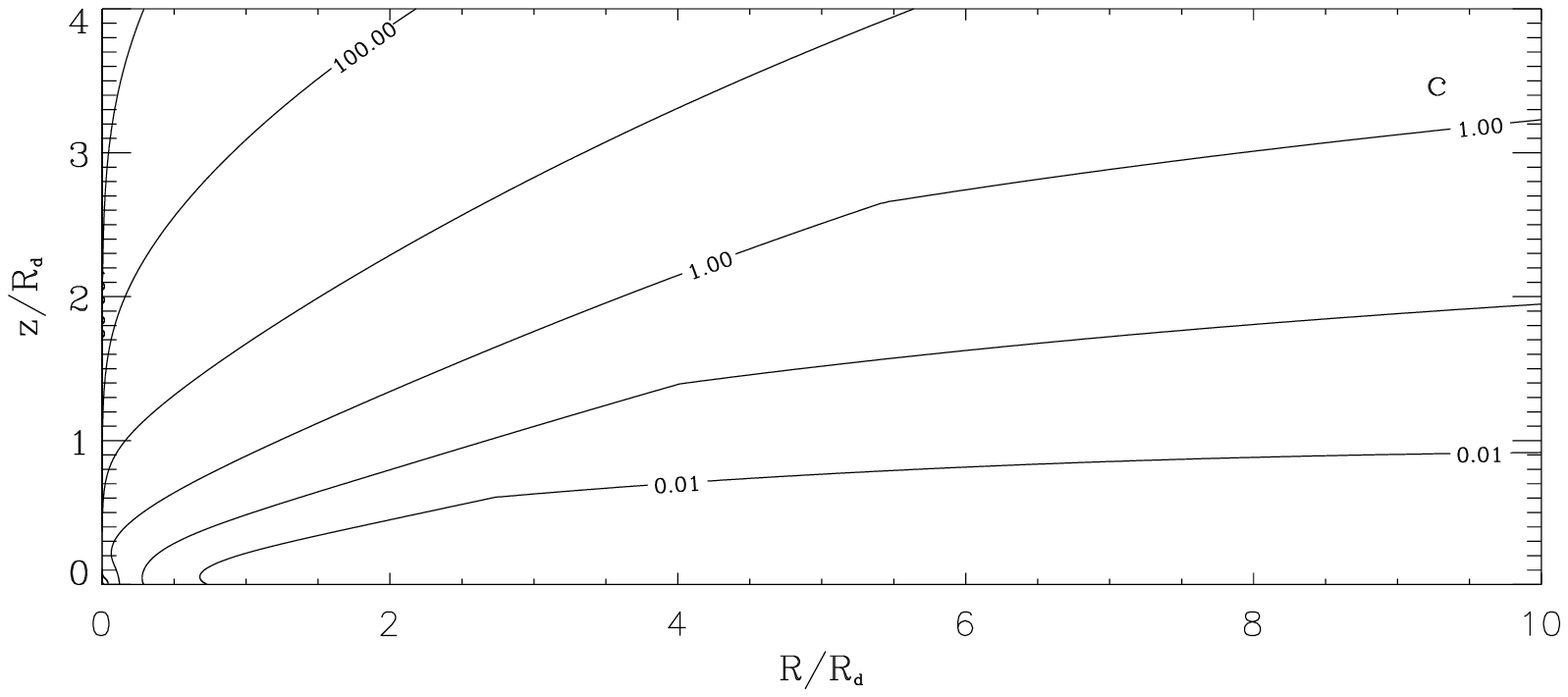}}
\caption{
\emph{Panel~a:} Bolometric
surface brightness isophotes (in erg s$^{-1}$ cm$^{-2}$) for the RM.
\emph{Panel~b:} meridional sections of equal cooling time (in
yrs) surfaces for the RM with $\rhong =2.4\times 10^{-26}$ g
cm$^{-3}$. \emph{Panel~c:} contours of equal $\xi = \tcool/\tdyn$
(Eq.~[\ref{xi}]) in the meridional plane.}
\label{isocool}
\end{figure}

Information on the thermal evolution of the RM is obtained from its
local cooling time 
\[ 
\tcool(R,z)\equiv {E\over\Ep}\propto{1\over\rhong}, 
\label{t.cool} 
\] 
where $E = 3 \rho k T / 2 \mu \mH$ and $\Ep = n_{\rm e} n_{\rm p}
\Lambda_0(T) $ are the gas internal energy and cooling rate per unit
volume; $k$, $\mu$, $\mH$, $n_{\rm e}$ and $n_{\rm p}$ are the
Boltzmann constant, the gas mean molecular weight, the proton mass,
and the number density of electrons and protons, respectively.  We
also adopt the cooling function (in erg cm$^3$ s$^{-1}$) 
\[
\Lambda_0(T)\simeq\left\{ \begin{array}{l} 5.36\times 10^{-27} T, \\
2.18\times 10^{-18} T^{-0.6826} + 2.71 \times 10^{-47} T^{2.976},
\end{array} \right.  
\label{Lambda} 
\] 
where the first equation holds for $10^4\leq T\leq 1.3\times 10^5$~K,
and the second for $1.3\times 10^5\leq T\leq 10^8$~K
(\citeauthor{Math.78} \citeyear{Math.78}, see also
\citeauthor{Ciot.91} \citeyear{Ciot.91}).  For simplicity we assume
$n_{\rm e} = n_{\rm p}$ ($\mu = 1/2$) and so 
\[
\Ep={\rho^2\over\mH^2}\Lambda_0 (T).  
\label{lum/vol.2} 
\] 

Note that, at variance with $\vel$ and $T$, $\Ep$ and $\tcool$ do
depend on the value of the gas density normalization constant $\rhong$
in Eq.~(\ref{rho.gas}), and so we are now forced to fix its value. A
simple argument can be based on the requirement of stationarity: the
gas distribution must radiate per unit time a total amount of energy
of the same order of magnitude as that provided by supernova
explosions and stellar winds in the stellar disks of NGC 891,
estimated $\approx 2.9 \times 10^{42}$ erg s$^{-1}$ \citep[ and
references therein]{Breg.97}. The total (bolometric) luminosity of the
gas distribution is 
\[ 
\Lt =4\pi\int_0^{\infty}\!\int_0^{\infty}\Ep R
\, {\rm d} R \, {\rm d}z \propto\rhong^2 , 
\label{totLum} 
\] 
and from numerical integration we find that $\rhong\approx 2.4\times
10^{-26}$ g cm$^{-3}$, corresponding to a total gas mass $M_{\rm
gas}\approx 1.6\times 10^9M_{\odot}$ (it should be noted that use
of $n_{\rm e}=1.2n_{\rm p}$ to include helium ionization would lead
to a density normalization and to a total mass $\approx$ 10\% smaller
than the values obtained here). The hypothesis of a negligible gas
self-gravity on galactic scale is thus satisfied, being the resulting
gas mass only a few percent of the total mass of the galaxy.
Curiously, the mass of the extra-planar gas distributed according to
(\ref{rho.gas}) ($\approx 5.5\times 10^8M_{\odot}$ for $z\gtrsim 1.4$
kpc), is found to be of the same order of magnitude as the mass of the
observed extra-planar \HI~\citep[$\approx 6\times 10^8
M_{\odot}$,][]{Swat.97}, while the luminosity of the RM extra-planar
gas is very low, summing up to only $\approx 5\%$ of the estimated
energy injection rate. From this point of view, the bulk of the
supernova and stellar wind heating is radiated by the gas near the
galactic disk, in good quantitative agreement with the
conclusions of \citet{Read.95}. This is confirmed by the
surface brightness distribution $\Sigma$ of the RM, obtained
projecting $\Ep(R,z)$ along the line-of-sight.  When the galaxy is
seen edge-on, the surface brightness is given by:
\[ 
\SB (R,z) = 2\int_R^{\infty} \frac{\Ep \, R' \, {\rm d} R'}
{\sqrt{{R'}^2 - R^2}} ,
\label{surfbr} 
\] 

In Fig.~\ref{isocool}a we show the RM bolometric surface brightness
distribution.  We have also computed the RM luminosity for the gas
with $T \geq 5.5 \times 10^{5}$~K (that we identify with EUV-emitting
gas), obtaining $L_{\rm EUV} \sim 10^{-2} \Lt$, with the bulk of the
emission confined in a hot bubble along the galactic axis and matching
the temperature distribution in Fig.~\ref{isorot}c.

The conclusion that a major fraction of the galaxy heating is radiated
near the disk is also confirmed by the fact that cooling times
increase from the galactic plane toward regions of higher~$z$ and
lower gas density, with values in the range
$10^6\lesssim\tcool\lesssim 10^{11}$ yrs
(Fig.~\ref{isocool}b). Obviously, the gas in the regions where
$\tcool$ is short will not remain homogeneous and its natural fate
will be to cool and form clouds and filaments.  The energy input from
the disk would be then used mainly to maintain a time-average,
stationary multiphase structure.

Additional insight on this multiphase medium is provided by the 
dynamical time
\[ 
\tdyn(R,z)\equiv {2\pi R\over\vel(R,z)},
\label{t.dyn}
\]
i.e., the orbital period of a gas element in circular orbit at
$(R,z)$.  It is natural to introduce here the dimensionless number
\[ 
\xi\equiv {\tcool\over\tdyn}\propto {1\over\rhong}.
\label{xi}
\]

From Fig.~\ref{isocool}c it appears that $\xi\gtrsim 1$ for $z\gtrsim
1\div 2\,\Rd$ and near the galaxy axis. It is expected that where $\xi
< 1$ the gas will cool locally. Note that the regions where $\xi <1$
match very closely those in which $\tcool <1$ Gyr.

In order to address the issue of the interaction of \HI~clouds with
the hot and homogeneous baroclinic gas distribution of the RM, we have
computed the cloud drag time $\tdrag$, i.e. the time required for an
\HI~cloud to corotate with the homogeneous gas halo described by
Eq.~(\ref{rho.gas}). This quantity is relevant whatever the origin of
the \HI~clouds: thermal instabilities of the hot gas, galactic
fountains or cosmological accretion.  Following \citet*{Klei.94}, we
define
\[
\tdrag (R,z)\equiv {8\over 3 C_{\rm D}}{\rc\over v_{\rm rel}}\chi
\propto {1\over\rhong} ,
\label{tdrag}
\]
where $C_{\rm D} \simeq 1$ is a numerical coefficient, $\rc$ is the
radius of a typical \HI~cloud, $v_{\rm rel}$ is the modulus of the
relative velocity between the cloud and the homogeneous extra-planar
component, and $\chi\equiv\rho_{\rm c}/\rho$ is the ratio between the
cloud and the medium densities.

We estimate $\tdrag$ assuming pressure equilibrium between cold and
hot gas components, from which $\chi =T/T_{\rm c}\approx 3000$.  A
fiducial value of the relative velocity is obtained by assuming
$v_{\rm rel}=\vel\approx 2\times 10^7$ cm s$^{-1}$, while the cloud
radius is estimated as $4\pi\rc^3\rho_{\rm c}/3 = 10^5
M_{\odot}\tilde{M}_5$, where $\tilde{M}_5$ is the cloud mass in units
of $10^5 M_{\odot}$. We thus obtain
\begin{eqnarray}
\tdrag &\simeq& {8\over 3\vel}\left( 
                {3\times 10^5 M_{\odot} \tilde{M}_5\over
                4\pi n_{\rm p}m_{\rm p}} 
                \right)^{1/3} 
                \left( 
                {T\over T_{\rm c}}\right)^{2/3}\approx\nonumber\\
        &&      2.7 \times 10^8\left( 
                {\tilde{M}_5\over n_{\rm p}} 
                \right)^{1/3}\textrm{yrs},
\label{tdrag2}
\end{eqnarray}
where $\rho =n_{\rm p}m_{\rm p}$.  From this simple formula it is
apparent that massive clouds at high $z$ above the galactic plane
cannot be dragged by the hot and tenuous RM distribution (see
Fig.~\ref{isorot}a for characteristic values of $n_{\rm p}$), while
smaller clouds (e.g.  $\tilde{M}_5\sim 10^{-3}$) near the disk
($z\approx 1\div 2$ kpc) have drag times significantly shorter than
rotational times. It cannot be excluded, therefore, that extraplanar
\HI~clouds (or, even better, \HI~filaments, which have larger
cross-sections) could be the tracers of the underlying fluid. Of
course, if \HI~clouds originated from cosmological accretion, then the
proper characteristic time to be compared with $\tdrag$ is the
free-fall time. However, $\tdyn$ is of the same order of magnitude and
the above arguments should still be valid. In any case, only
hydrodynamical simulations will possibly clarify the importance of
drag and related issues, such as the angular momentum redistribution
between the homogeneous gas and the \HI~clouds.  In conclusion, our
analysis seems to leave open the possibility advocated by Collins et
al. of the presence of a non-gravitational forces acting on the
extra-planar \HI~gas.

\subsubsection{A ``gas'' of \HI~clouds}
\label{int:clouds}

As pointed out at the beginning of Sect.~\ref{ngc891}, from a formal
point of view, Eqs.~(\ref{eq.hydro}) are identical to the
fluid stationary Jeans equations for an axisymmetric system with a
globally isotropic velocity dispersion tensor. Thus, also for such a
system we would obtain rotational velocities decreasing with
increasing $z$. The pressure field, however, must be interpreted as
$P=\rho\sigma^2$, where $\sigma$ is the clouds (1-dimensional)
velocity dispersion, so that $\sigma=\sqrt{k T/\mu m_{\rm p}}\approx
41\sqrt{\tilde T_5}$ km s$^{-1}$, where $\tilde T_5\equiv T/10^5$ K.
From Fig.~\ref{isorot}c it appears that near the disk the clouds
velocity dispersion would be of the order of 30 km s$^{-1}$ or even
less, while higher values of $\sigma$ are expected near the galaxy
center. Such values might not be unreasonable. However, a further
comparison of the hot and homogeneous gas solution with the present
one is not straightforward: while the RM gas distribution in
Eq.~(\ref{rho.gas}) was adopted to describe a hot and homogeneous gas
distribution (and thus not very well constrained by observations), a
proper Jeans-based analysis would require the choice of a cloud
distribution similar to that observed. In principle this should be
simple to do, but we do not explore this any further here. At any
rate, even if the temperature and related problems addressed in
Sect.~\ref{int:gas} were solved, in the Jeans-based interpretation the
question of how the clouds velocity dispersion tensor is rendered
isotropic would become the new central question.

\section{Discussion and conclusions}
\label{concl}

We have presented a family of fluid stationary models for the
extra-planar gas in spiral galaxies. As an application, we have built
a model for the extra-planar gas of the spiral galaxy NGC~891. The
main results of our analysis are:

\begin{itemize}

\item[(i)] Physically acceptable baroclinic solutions exist for gas 
density distributions (isodensity surfaces) more flattened than the
isopotential surfaces. For regions near the disk, the condition is
that the gas density distribution is centrally depressed.

\item[(ii)] Application of our method to the case of the edge-on 
galaxy NGC 891 has shown that with baroclinic solutions it is possible
to reproduce the observed vertical decrease of the gas rotational
velocity.

\item[(iii)] In homogeneous fluid stationary solutions the gas 
temperature is in the range $10^4\lesssim T \lesssim 10^6$~K, well
above the \HI~temperature and the cooling times are short, of the
order of the orbital times.

\item[(iv)] In the hot, homogeneous gas configurations with 
negative velocity gradients of the present models, the drag on small
\HI~clouds ($\sim10^2 M_{\odot}$) near the disk is important and may 
account for the observed vertical velocity decrease. It is not
certain, however, that this applies to the case of NGC 891 where the
\HI~structures may be more massive.

\item[(v)] Instead of a hot homogeneous medium, the baroclinic
solution can also provide a good fluid description of a ``gas'' of
clouds (according to the stationary Jeans equations) with a globally
isotropic velocity dispersion tensor. All the above considerations
about the decrease of rotational velocity with increasing $z$ would
still hold, while the temperature field would have to be interpreted
as the velocity dispersion field.

\end{itemize}

Ballistic models have been tried for the extra-planar gas.  It seems,
however, that in order to reproduce the observed rotational velocity
gradients, also non-gravitational effects, such as gas and magnetic
pressure, may be necessary \citep{Coll.02, Frat.05b}. Our analysis
seems to suggest that such non gravitational effects could be due (at
least for low mass clouds) to ram pressure.  Note that $\tdrag$
depends on the \emph{relative} velocity between the \HI~clouds and the
hot baroclinic gas, hence the effect of the drag would be to
\emph{regularize} the motion of the clouds to the velocity field of
the embedding medium. For example, if the \HI~clouds are ejected from
the disk by galactic fountains, they will be, on average, initially
faster with respect to the baroclinic gas and will be slowed down to
its velocity while rising.  Incidentally, the drag could also be
effective in regularizing the dynamics of accreted clouds: from this
point of view, a regular \HI~dynamics (in particular, the same sense
of rotation of the galactic disk) would not be necessarily an
indication of internal origin for the extra-planar gas.

The problem of stability of the models presented here is a
difficult one and remains open.  Numerical hydrodynamical simulations
would be very useful to address it. In the approach adopted in this
paper, where the density distribution for the extraplanar gas is
assigned, pressure, temperature and rotation cannot be arbitrarily
prescribed but are determined by the galaxy gravitational
potential. The stability of such configurations is not guaranteed.
Additional effects worth studying and needing quantitative estimates
are the drag of clouds (both of internal and external origin), their
cooling and evaporation times, and the overall energy budget, to
follow up on the pioneering work of \citet{Cox.74} and
\citet{McKee.77}. Equally important would be the numerical study of
baroclinic solutions in the context of Jeans equations. In this case,
isotropy of the velocity dispersion tensor could be provided by
cloud-cloud collisions, a physical ingredient usually neglected in
numerical simulations \citep[however, see][]{Waxm.78}. One could
speculate that the problems encountered by purely ballistic models is
due to the absence of a collision term in the simulations, as already
suggested by \citet{Coll.02}.

We conclude remarking that fluid-homogeneous and Jeans-based
baroclinic models do not exclude each other. Perhaps, they could be 
integrated and used together for a better understanding of the 
dynamics of the extra-planar gas.

\begin{acknowledgements} 

We thank an anonymous referee for insightful comments.  We are
grateful to Giuseppe Bertin, James Binney, Jeremiah Ostriker, and Hugo
van Woerden for enlightening discussions. We also thank Fabrizio
Brighenti and Annibale D'Ercole for useful comments.  M.B. was
partially supported by ASI contract IR/063/02 and by INAF-Bologna
Astronomical Observatory. L.C. and R.S. were supported by the grant
CoFin2004 (MIUR).

\end{acknowledgements}

\appendix

\section{A simple analytical toy-model}
\label{toymodel}

We present here a simple and fully analytical toy-model of a
baroclinic gas distribution characterized by the vertical decline of
the rotational velocity.  The dimensionless gas density distribution 
\[
\tilde\rho = {1-\etag\lambda\over {\tilde r}^{\lambda}}+
             {\etag\lambda\Rti^2\over {\tilde r}^{2+\lambda}} 
                               ,\quad (0<\lambda <3),
\label{toy:rhogas.hom}
\]
is obtained from a homeoidal expansion of the oblate power-law
distribution $\rho_0/m^{\lambda}$, where $\tilde\rho\equiv
\rho/\rho_0$, $r=\sqrt{R^2+z^2}$, $m^2=\Rti^2 +{\tilde
z}^2/(1-\etag)^2$, $\Rti\equiv R/\Rg$, $\tilde z\equiv z/\Rg$, and
$0\leq\etag\leq 1/\lambda$ is the flattening of the distribution
\citep{Ciot.05}.

The galaxy (stars plus dark matter) density distribution is also
described by the homeoidal expansion of $\rhonda/m^{\gamma}$, (where
$0<\gamma <3$, $\rhonda$ is a normalization density, $m$ is defined
like above but with $\etah$ replacing $\etag$, and $0\leq\etah\leq
1/\gamma$). Without loss of generality also the galaxy scale-length is
$\Rg$, and the gravitational potential for $\gamma\neq 2$ (in units of
$4\pi G\Rg^2\rhonda$) is given by
\[
\tilde\Phi = -{5-\gamma -\etah (4-\gamma)(\gamma -1)\over
              (5-\gamma)(3-\gamma)(\gamma-2){\tilde r}^{\gamma-2}}-
              {\etah\Rti^2\over (5-\gamma){\tilde r}^{\gamma}}
\label{toy:PhiDM}
\]
(Ciotti \& Bertin, Eq. [27]).  The pressure (normalized to $4\pi
G\Rg^2\rhonda\rho_0$), and the square of the gas rotational velocity
field (normalized to $4\pi G\Rg^2\rhonda$) are obtained from
Eqs.~(\ref{pressione})-(\ref{velocita}).  For simplicity we report
here the solutions up to the linear terms in the flattenings, even
though the derivation of the full solutions (i.e., including also the
terms in $\etah\etag$) presents no difficulty. Accordingly, 
\begin{eqnarray} 
\tilde P &=&{{\tilde r}^{2-\lambda-\gamma}\over (5-\gamma)(3-\gamma)}
          \Bigg[
          {(5-\gamma)(1-\etag\lambda)-(4-\gamma)(\gamma-1)\etah\over 
          \gamma+\lambda-2}\nonumber\\
          &&+{(3-\gamma)\gamma\etah +(5-\gamma)\etag\lambda\over 
              \gamma+\lambda}{\Rti^2\over {\tilde r}^2}
          \Bigg],
\label{toy:pre}
\end{eqnarray}
\[ 
\tilde\rho\tilde\velq ={2\lambda\Rti^2\over 
                        (\gamma+\lambda){\tilde r}^{\gamma +\lambda}}
                       \left ( 
                       {\etag\over 3-\gamma}-{\etah\over 5-\gamma}\right).
\label{toy:velq}
\]

\begin{figure}[!thb] 
\resizebox{0.95\hsize}{!}{\includegraphics{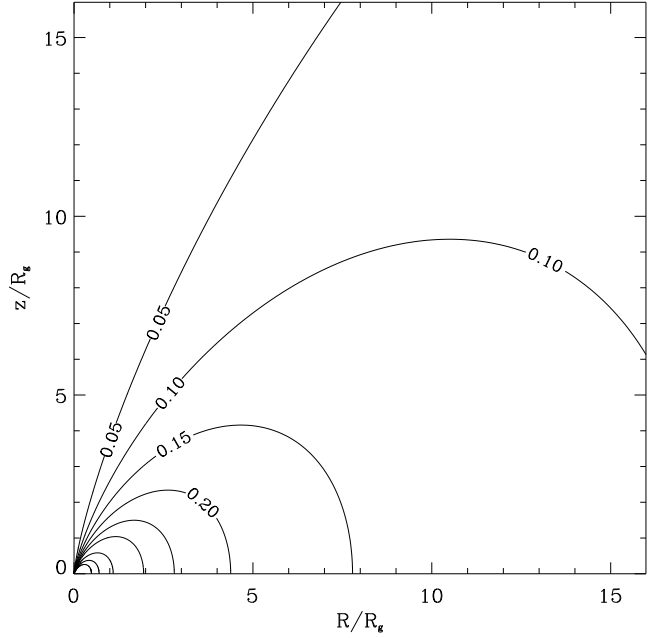}}
\caption{Meridional section of the isorotation surfaces $\tilde\vel^2$ 
         (labels along the curves) for $\gamma = 2.5$, $\etah = 0.2$, 
         $\lambda =2.3$ and $\etag = 0.3$.}
\label{isorot.toy}
\end{figure}

Note that, for $\etag < \etah (3-\gamma)/(5-\gamma)$ the (linearized)
solution is unphysical. The isorotation curves for a physically
acceptable case (obtained from the full solution) are shown in
Fig~\ref{isorot.toy}, where it is apparent how, even in this extremely
simplified model, the rotational velocity decreases with increasing
$z$.

\section{How to construct $\rhope$}
\label{buildrho}

A first approach to the construction of the distribution $\rhope$
considered in Sect.~\ref{factor.rho} is the assumption that $\pp
=\pp(\rho)$, as for example $\pp\propto \rho^{\gamma}$.  In this case,
from Eq.~(\ref{hydroeq.eff})
\[ 
\rhope = \rhong \times \left\{
\begin{array}{l}
   \displaystyle
   \exp \left[ -\frac{\mu \mH}{k T_0} 
   (\phieff - \Phi_{\rm eff, 0}) \right] , \\
   \\
   \displaystyle
   \left[ 1 + \frac{\gamma - 1}{\gamma}
   \, \frac{\mu \mH}{k T_0} \, (\Phi_{\rm eff, 0} - \phieff) 
   \right]^{\frac{1}{\gamma - 1}} ,
\end{array}
\right.
\label{rho.baro}
\]
where $\rhong$ and $\Phi_{\rm eff, 0}$ are taken at the same arbitrary
but fixed point $(R_0, z_0)$, and the first expression holds for the
isothermal case.  In the second case $T/T_0 = (\rhoe/\rhong)^{\gamma
-1}$ and, at variance with the isothermal case, a truncation of
$\rhoe$ may appear.  A different approach to the construction of
$\rhoe$ is however possible, where the specific density field is
prescribed. This can be done by using the well known property that
solutions of Eq.~(\ref{hydroeq.eff}) necessarily are stratified on
$\phieff$, i.e. $\rhoe =
\rhope$ and $\pp =\pp (\phieff)$.  Thus, if one fixes $\phieff$ and a 
prescribed function $\rhope$, the pressure field is obtained by direct
integration as
\[
\pp (\phieff)=\pp (\Phi_{\rm eff, 0})- 
              \int_{\Phi_{\rm eff, 0}}^{\phieff}\rhoe(t) \, {\rm d}t.
\label{P(phieff)} 
\]

Of course, this approach can be used only for density stratifications
such that $\pp > 0$ everywhere.

\section{The numerical code}
\label{codice}

We describe here the scheme on which the (double precision) FORTRAN77
code adopted for the construction of models in Sect.~\ref{ngc891} is
based.  From symmetry arguments we restrict the computation to the
region $\left[\Rmin, \Rmax \right] \times \left[\zmin, \zmax \right]$
in the half-space $z \geq 0$, on which we define a bilogarithmic grid
of $(n_R + 1) \times (n_z + 1)$ elements.  In the radial direction the
grid is
\[
\Ri=R_1\times 10^{(i-1)\Delta R},
\]
where $R_0=0$, $R_1=\Rmin$, and 
\[
\Delta R\equiv {\log\Rmax -\log\Rmin\over n_R -1} ,
\]
so that $\Rmax=R_{n_R}$; the $z$ coordinate is discretized in the same
way.

The gas density and the gravitational field components at $(\Ri ,\zj)$
are $\rhoij$, $\phiRij\equiv (\partial\Phi/\partial R)_{ij}$ and
$\phizij\equiv (\partial\Phi/\partial z)_{ij}$.  For fixed $\Ri$ we
integrate the first of Eqs.~(\ref{eq.hydro}) with linear interpolation
and boundary condition $\pp (\Ri,\zmax) = 0$:
\[
\ppij ={1\over 2} \sum_{k=j}^{n_z -1}
       \Delta z_k\times 
       (\rho^i_k {_z\Phi^i_k}+\rho^i_{k+1} {_z\Phi^i_{k+1}}) ,
\label{f77:pres}
\]
where $\ppij \equiv \pp(\Ri ,\zj)$ and $\Delta z_k\equiv z_{k+1}-z_k$.
Accordingly, from Eq.~(\ref{velocita}):
\[
\rhoij (\velq)^i_j = (\ppRij + \phiRij\rhoij)\Ri,
\label{f77:vel}
\]
where $\ppRij \equiv (\pp^{i+1}_j - \ppij)/\delRi$ and $\delRi
\equiv\Rii -\Ri$. The gas temperature $T$, the cooling rate $\Ep$
(Eq.~[\ref{lum/vol.2}]), the cooling time $\tcool$
(Eq.~[\ref{t.cool}]) and the ratio $\xi$ between cooling time and
dynamical time (Eq.~[\ref{xi}]) are then obtained on the grid points.

The gas total luminosity is obtained as 
\[
\Lt =\sum_{i=0}^{n_R-1}\sum_{j=0}^{n_z-1}L^i_j,
\]
where 
\[
L^i_j= 4\pi\int_{\Ri}^{\Rii}\int_{\zj}^{\zjj} R 
\, \Ep(R,z) \, {\rm d}R \, {\rm d}z,
\label{Lij}
\]
and $\Ep(R,z)$ is evaluated on the numerical grid from
Eq.~(\ref{lum/vol.2}). The integral in Eq.~(\ref{Lij}) is evaluated
analitically on each region $[\Ri, \Rii]\times [\zj, \zjj]$ by
considering the bilinear expansion
\[ 
\Ep (R,z) = A_0 + A_1 R + A_2 z + A_3 Rz ,
\label{bilinear}
\]
where
\begin{eqnarray*}
A_0 & \equiv & \epa - \frac{\epb-\epa}{\delRi} \Ri - 
               \frac{\epc-\epa}{\delzj} \zj + \\
    &        & \frac{\epd-\epb+\epa-\epc}{\delRi \delzj} \Ri \zj \\
A_1 & \equiv & \frac{\epb-\epa}{\delRi} - 
	\frac{\epd-\epb+\epa-\epc}{\delRi \delzj} \zj \\
A_2 & \equiv & \frac{\epc-\epa}{\delzj} - 
	\frac{\epd-\epb+\epa-\epc}{\delRi \delzj} \Ri \\
A_3 & \equiv & \frac{\epd-\epb+\epa-\epc}{\delRi \delzj} .
\end{eqnarray*}

With this formula,
\begin{eqnarray}
L^i_j &=& \pi\delRi\delzj\Big[2A_0 (\Rii +\Ri)+   \nonumber\\ 
      &&  {4A_1\over 3} (\Rii^2 +\Rii\Ri +\Ri^2)+ \nonumber\\
      &&  A_2 (\Rii +\Ri)(\zjj +\zj)+             \nonumber\\
      &&  {2A_3\over 3}(\Rii^2 +\Rii\Ri +\Ri^2)(\zjj +\zj)\Big].
\end{eqnarray}
Finally, the edge-on surface brightness $\Sigma$ is given by the
discretization of Eq.~(\ref{surfbr}) and so,
in our scheme, 
\[
\Sigma^i_j =\sum_{k=i}^{n_R -1} 
            (\Ep^{k+1}_j +\Ep^k_j)
            \left (\sqrt{R_{k+1}^2-\Ri^2}-\sqrt{R_k^2 -\Ri^2}\right).
\label{code:surfbr.d}
\]

\end{document}